\documentclass[aps,twocolumn,floatfix,superscriptaddress]{revtex4-1}

\usepackage{bm}
\usepackage{amsmath,amssymb,cases}
\usepackage{braket}
\usepackage{graphicx}
\usepackage{color}

\usepackage{ulem}

\begin{document}

\title{Systematic study on the dependence of the warm-start quantum approximate optimization algorithm on approximate solutions}

\author{Ken N. Okada}
\affiliation{Center for Quantum Information and Quantum Biology, Osaka University, Osaka 560-8531, Japan}
\author{Hirofumi Nishi}
\author{Taichi Kosugi}
\affiliation{Laboratory for Materials and Structures, Institute of Innovative Research, Tokyo Institute of Technology, Tokyo, 152-8550, Japan}
\affiliation{Quemix Inc., Tokyo, 103-0027, Japan}
\author{Yu-ichiro Matsushita}
\affiliation{Laboratory for Materials and Structures, Institute of Innovative Research, Tokyo Institute of Technology, Tokyo, 152-8550, Japan}
\affiliation{Quemix Inc., Tokyo, 103-0027, Japan}
\affiliation{Quantum Material and Applications Research Center, National Institutes for Quantum Science and Technology, Tokyo, 152-8552, Japan}

\begin{abstract}
Quantum approximate optimization algorithm (QAOA) is a promising hybrid quantum-classical algorithm to solve combinatorial optimization problems in the era of noisy intermediate-scale quantum computers. 
Recently warm-start approaches have been proposed to improve the performance of QAOA, where approximate solutions are obtained by classical algorithms in advance and incorporated into the initial state and/or unitary ansatz.
In this work, we study in detail how the accuracy of approximate solutions affect the performance of the warm-start QAOA (WS-QAOA). 
We numerically find that in typical MAX-CUT problems, WS-QAOA tends to outperform QAOA as approximate solutions become closer to the exact solutions in terms of the Hamming distance.
We reveal that this could be quantitatively attributed to the initial state of the ansatz. 
We also solve MAX-CUT problems by WS-QAOA with approximate solutions obtained via QAOA, having a better result than QAOA especially when the circuit is relatively shallow.
We believe that our study may deepen understanding of the performance of WS-QAOA and also provide a guide as to the necessary quality of approximate solutions. 
\end{abstract}

\maketitle

\section{Introduction}\label{sec:intro}
The last decade has seen significant technological progress in manufacturing hardware platform of quantum computers~\cite{Kjaergaard2020}. 
The current pace of scale-up in quantum devices raises a hope that quantum processors with hundreds of physical qubits could be available within the next decade.
These near-term quantum computers are referred to as noisy intermediate-scale quantum (NISQ) computers~\cite{Preskill2018} in that they are classically intractable, but still not sufficiently large to implement quantum error correction. 
As the NISQ era approaches, there have been an increasing number of researches that develop algorithms to efficiently leverage NISQ devices~\cite{McArdle2020, Cerezo2021, bharti2021}. They are designed to solve quantum many-body problems in chemistry and physics as well as classical problems in combinatorial optimization and machine learning. 
Most of these studies employ hybrid quantum-classical approaches, primarily variational quantum algorithms~\cite{Peruzzo2014, McClean2016}.
In these algorithms, variational quantum states are created via parameterized quantum circuits on a quantum computer, whereas the parameters are updated on a classical computer to optimize the objective function calculated with the measurement outcomes. 
Since variational quantum algorithms generally take a relatively low number of gate operations, they are considered as suitable to gain quantum advantage on NISQ computers.

Quantum approximate optimization algorithm (QAOA)~\cite{Farhi2014}, a representative example of variational quantum algorithms, solves combinatorial optimization problems in a spirit analogous to adiabatic quantum annealing (QA)~\cite{Das2008, Albash2018, Hauke2020}. 
Indeed, the variational state (ansatz) of QAOA can be deduced by the Trotter decomposition of the time evolution of QA.
Despite some numerical demonstrations of its efficacy in small-size problems~\cite{Crooks2018, Zhou2020}, it has been a subject of discussions whether QAOA could practically outperform the best classical algorithms~\cite{Hastings2019, Bravyi2020}.
Several kinds of variants have been proposed to improve upon the original version of QAOA~\cite{Farhi2017, Zhu2017, Bartschi2020, Hadfield2019, Wang2020, Egger2021, Tate2020}.
To name a few, Farhi~et.~al proposed a variant of the ansatz by allowing different parameters for each rotation gate~\cite{Farhi2017}.
Hadfield~et.~al extended the ansatz by generalizing mixer operations, which could be suitable to optimization problems with constraints~\cite{Hadfield2019}.

The variant that our work will focus on is the warm-start QAOA (WS-QAOA) proposed by Egger~et.~al~\cite{Egger2021} and Tate~et.~al~\cite{Tate2020}.
The basic idea behind the approach is to facilitate the convergence to the solution by distorting the original ansatz towards a classically-obtained approximate solution.
Egger~et.~al encodes rounded/unrounded semidefinite programming solutions into the initial state and mixer term in the ansatz~\cite{Egger2021}. 
Tate~et.~al also encodes semidefinite programming relaxations into the initial state, but not the unitary circuit~\cite{Tate2020}. 
We also mention that a similar warm-start approach has been independently studied in QA~\cite{Grass2019}.  

In this paper, we examine how the performance of WS-QAOA depends on quality of approximate solutions to make a deep understanding of its efficacy.
In refs.~\cite{Egger2021, Tate2020}, the authors solved problems with WS-QAOA by acquiring approximate solutions by classical algorithms. Meanwhile, it remains unclear how accurate approximate solutions should be for WS-QAOA to outperform QAOA.    
Here we deduce the ansatz of WS-QAOA starting from QA with a bias field~\cite{Grass2019} and carefully study how the performance of WS-QAOA depends on the quality of approximate solutions by numerical simulations on the MAX-CUT problem. 
We find out that WS-QAOA shows a better performance relative to QAOA as the Hamming distance of the approximate solutions to the exact ones becomes smaller. 
We also reveal that the observation could be partially attributed to the initial state of the ansatz.
Finally, we solve the MAX-CUT problem with WS-QAOA after obtaining approximate solutions by QAOA and have superior results to QAOA especially when the circuit depth is small.

The rest of the paper is organized as follows.
In Sec.~\ref{sec:formulation}, we formulate WS-QAOA in the context of QA.
Then, in Sec.~\ref{sec:WS-QAOA_numerics}, we numerically study the performance of WS-QAOA on the MAX-CUT problem for various approximate solutions in terms of the Hamming distance to the exact solutions as well as for different strengths of the bias field.
In Sec.~\ref{sec:QAOA+WS}, we solve the MAX-CUT problem by combining WS-QAOA with QAOA and compare its efficacy to QAOA.
Finally, in Sec.~\ref{sec:conclusion}, we summarize our results.

\section{Formulation}
\label{sec:formulation}
\subsection{MAX-CUT problem}
As a prototypical combinatorial optimization problem, we consider the MAX-CUT problem, which is known as NP-hard.
It is defined on a graph $G=(V, E)$, where $V$ represents a set of vertices, and $E$ represents a set of edges between the vertices.
We denote the number of vertices in $G$ as $n$. 
The MAX-CUT problem is to find a partition of $V$ into two subsets that maximizes the total number of edges between one subset and the other.
In a general case that each edge is associated with a real-valued weight $w_{ij}$, one evaluates the weighted sum of those edges. 
The problem is formulated as maximization of the following objective function 
\begin{equation} 
C(\{x_i\})=\sum_{(i,j)\in E}w_{ij}(x_i(1-x_j)+x_j(1-x_i)),
\label{eq:obj_function}
\end{equation}
where $x_i$ denotes a binary variable associated with vertex $i$ $(x_i=0, 1)$.
We note that bit strings $\{x_i\}$ and $\{\overline{x}_i\}$ ($\overline{x}_i\equiv 1-x_i$) give the same value of $C$.
In the following, we denote $\{x^{\rm sol}_i\}$ and $\{\overline{x}^{\rm sol}_i\}$ as the single pair of solutions.

In the language of physics, the MAX-CUT problem is encoded in finding the ground state of the corresponding Ising Hamiltonian, which is obtained by replacing $x_i$ for $(1-Z_i)/2$ ($Z$: the Pauli $Z$ matrix) in the objective function $C$ and changing the whole sign. The Hamiltonian reads 
\begin{equation} 
H_{\rm C}=\sum_{(i,j)\in E}\frac{w_{ij}}{2}Z_iZ_j,
\label{eq:H_obj}
\end{equation}
where the offset 
\begin{equation} 
D=-\sum_{(i,j)\in E}\frac{w_{ij}}{2}
\label{eq:offset}
\end{equation}
is left off. 

\subsection{QAOA}
QAOA searches for the ground state of the Hamiltonian $H_{\rm C}$ using a QA-inspired ansatz with $2p$ variational parameters for depth $p$~\cite{Farhi2014}.
The ansatz is constructed by alternating applications of the driver operation $U_{\rm C}$ and mixer operation $U_{\rm T}$ to the equal-weight superposition state $\ket{+}^{\otimes n}$. It is written down with variational parameters $\beta_s$ and $\gamma_s$ $(1\leq s\leq p)$ as 
\begin{equation}
\ket{\Psi_{\rm QAOA}}=\prod_{s=1}^pU_{\rm T}(\beta_s)U_{\rm C}(\gamma_s)\ket{+}^{\otimes n}.
\label{eq:QAOA_ansatz}
\end{equation}
Here the driver and mixer are defined as $U_{\rm C}(\gamma_s)=e^{-i\gamma_s H_{\rm C}}$ and $U_{\rm T}(\beta_s)=e^{-i\beta_s H_{\rm T}}$, respectively, where $H_{\rm T}=-\sum_iX_i$ ($X$: the Pauli $X$ matrix) represents a transverse-field term.
One can deduce the ansatz $\ket{\Psi_{\rm QAOA}}$ via the first-order Trotter decomposition of the QA procedure, where the wave function evolves under the Hamiltonian
\begin{equation}
H_{\rm QA}(t)=(1-u(t))H_{\rm T}+u(t)H_{\rm C}
\label{eq:H_QA}
\end{equation}
with a schedule function $u(t)$ ($u(0)=0$ and $u(T)=1$).

\subsection{WS-QAOA}
QA has had considerable success in solving combinatorial optimization problems~\cite{Das2008, Albash2018, Hauke2020}. 
However, when gap closing occurs during the annealing, it often gets stuck at suboptimal solutions.
Reverse QA is an effective variant to circumvent this challenge, which incorporates in the annealing process an approximate solution obtained in advance~\cite{Perdomo-Ortiz2011}.
In this procedure, the state adiabatically evolves from the approximate solution at the beginning to the exact solution at the end, driven by quantum fluctuations of a transverse field with a mountain-like time profile. 
The dynamics is described by the Hamiltonian $H_{\rm RQA}(t)=(1-t/T)H_{\rm I}+h(t)H_{\rm T}+(t/T)H_{\rm C}$ $(0\leq t\leq T)$, where $H_{\rm I}$ yields the approximate solution as the ground state, and $h(t)$ is a concave function with $h(0)=h(T)=0$. 
It was shown that the performance of reverse QA is largely dominated by the Hamming distance of the approximate solution from the exact one~\cite{Perdomo-Ortiz2011}.

Recently Gra$\ss$~\cite{Grass2019} proposed a similar but simpler QA procedure to make use of an approximate solution, which introduces a longitudinal bias field that favors the approximate solution. 
The procedure, which we call biased quantum annealing (BQA) hereafter, is governed by the Hamiltonian 
\begin{equation}
H_{\rm BQA}(t)=(1-u(t))(H_{\rm T}+H_{\rm L})+u(t)H_{\rm C},
\label{eq:H_BQA}
\end{equation}
where $H_{\rm L}$ represents a site-dependent longitudinal field defined as
\begin{equation}
H_{\rm L}=-\alpha\sum_i\left(1-2x_i^0\right)Z_i.
\end{equation}
Here $\{x_i^0\}$ represents an approximate solution, and $\alpha$ denotes strength of the bias field. 
The author showed that BQA outperforms QA when one prepares approximate solutions that are close enough to the exact solutions in terms of the Hamming distance~\cite{Grass2019}.

Here we formulate a QAOA version of BQA, which actually corresponds to WS-QAOA~\cite{Egger2021, Tate2020}.
One can derive the ansatz via the Trotter decomposition of BQA under the Hamiltonian $H_{\rm BQA}(t)$ in the same manner as one deduces $\ket{\Psi_{\rm QAOA}}$ from $H_{\rm QA}(t)$. 
Then the WS-QAOA ansatz is represented as 
\begin{equation}
\ket{\Psi_{\rm WS-QAOA}}=\prod_{s=1}^pU_{\rm T}(\beta_s)U_{\rm L}(\beta_s)U_{\rm C}(\gamma_s)\ket{\Psi_0},
\label{eq:WS-QAOA_ansatz}
\end{equation}
where $U_{\rm L}$ is defined as $U_{\rm L}(\beta_s)=e^{-i\beta_s H_{\rm L}}$. The initial state $\ket{\Psi_0}$ is written down as 
\begin{equation}
\ket{\Psi_0}=\prod_{i=1}^n\otimes R_Y\left(-\frac{\pi}{2}+\left(1-2x_i^0\right)\tan^{-1}\alpha\right)\ket{0}.
\label{eq:WS-QAOA_initial_state}
\end{equation}
$R_Y(\theta)=e^{i(\theta/2)Y}$ ($Y$: the Pauli $Y$ matrix) represents a $\theta$-rotation around the $y$-axis.
For $\alpha=0$, $\ket{\Psi_{\rm WS-QAOA}}$ corresponds to the QAOA ansatz $\ket{\Psi_{\rm QAOA}}$.
The WS-QAOA ansatz $\ket{\Psi_{\rm WS-QAOA}}$ is almost identical to that in ref.~\cite{Egger2021} except a small difference in representation of the mixer; the latter implements $e^{-i\beta_s (H_{\rm T}+H_{\rm L})}$ with three layers of rotation gates, whereas the former uses a decomposed form $e^{-i\beta_s H_{\rm T}}e^{-i\beta_s H_{\rm L}}$ with two layers. 

\section{Numerical simulations}\label{sec:WS-QAOA_numerics}
In this section, we examine how the WS-QAOA performance varies with choice of approximate solutions $\{x_i^0\}$.
For that purpose, we numerically study the MAX-CUT problem on weighted 3-regular (w3R) graphs. 
In w3R graphs, each vertex is connected to three others chosen at random, and each edge has weight $w_{ij}$ randomly set from $[0,1)$. 
We employ a fast quantum circuit simulator Qulacs~\cite{Suzuki2020}. 

For optimization of the parameters, we use two methods, random initialization (RI) and an interpolation-based heuristic termed INTERP~\cite{Zhou2020}. 
In RI, we take the best sample out of 50 randomizations of the initial values.
Given the translational symmetry of the ansatz $\ket{\Psi_{\rm WS-QAOA}}$, initial values of $\beta_s$ are set from $[-\frac{\pi}{4}, \frac{\pi}{4})$ for $\alpha=0$, $[-\frac{\pi}{2}, \frac{\pi}{2})$ for $\alpha=1$, and $[-\pi, \pi)$ otherwise, whereas those of $\gamma_s$ are set from $[-2\pi, 2\pi)$.
On the other hand, in INTERP, the parameters are optimized incrementally from depth $1$ to depth $p$. 
Here initial values of the parameters at depth $p$, $\beta_s[p]$ and $\gamma_s[p]$ $(1\leq s\leq p)$, are uniquely determined via an interpolation of the optimized values at depth $p-1$ as $\beta_s[p]=\frac{s-1}{p-1}\beta_{s-1}[p-1]+\left(1-\frac{s-1}{p-1}\right)\beta_s[p-1]$ ($\beta_0[p-1]=\beta_p[p-1]=0$).
It has been revealed that INTERP works more efficiently than RI for QAOA on w3R graphs~\cite{Zhou2020}. 
In this work, based on our benchmark calculations, we choose a better method, depending on $\alpha$, $p$. For QAOA ($\alpha=0$), we use INTERP. For WS-QAOA, with $\alpha=0.4$, we use INTERP, whereas, with $\alpha=1$, we use RI at $p\leq 3$ and INTERP at $p=4$. 
We note that, regardless of $\alpha(\neq0)$, we use RI when $\{x_i^0\}$ corresponds to the exact solution.
In both methods, parameters are updated via a gradient descent until the gradient becomes lower than a certain threshold value.

\begin{figure}[htb]
\begin{center}
\includegraphics[width=\columnwidth]{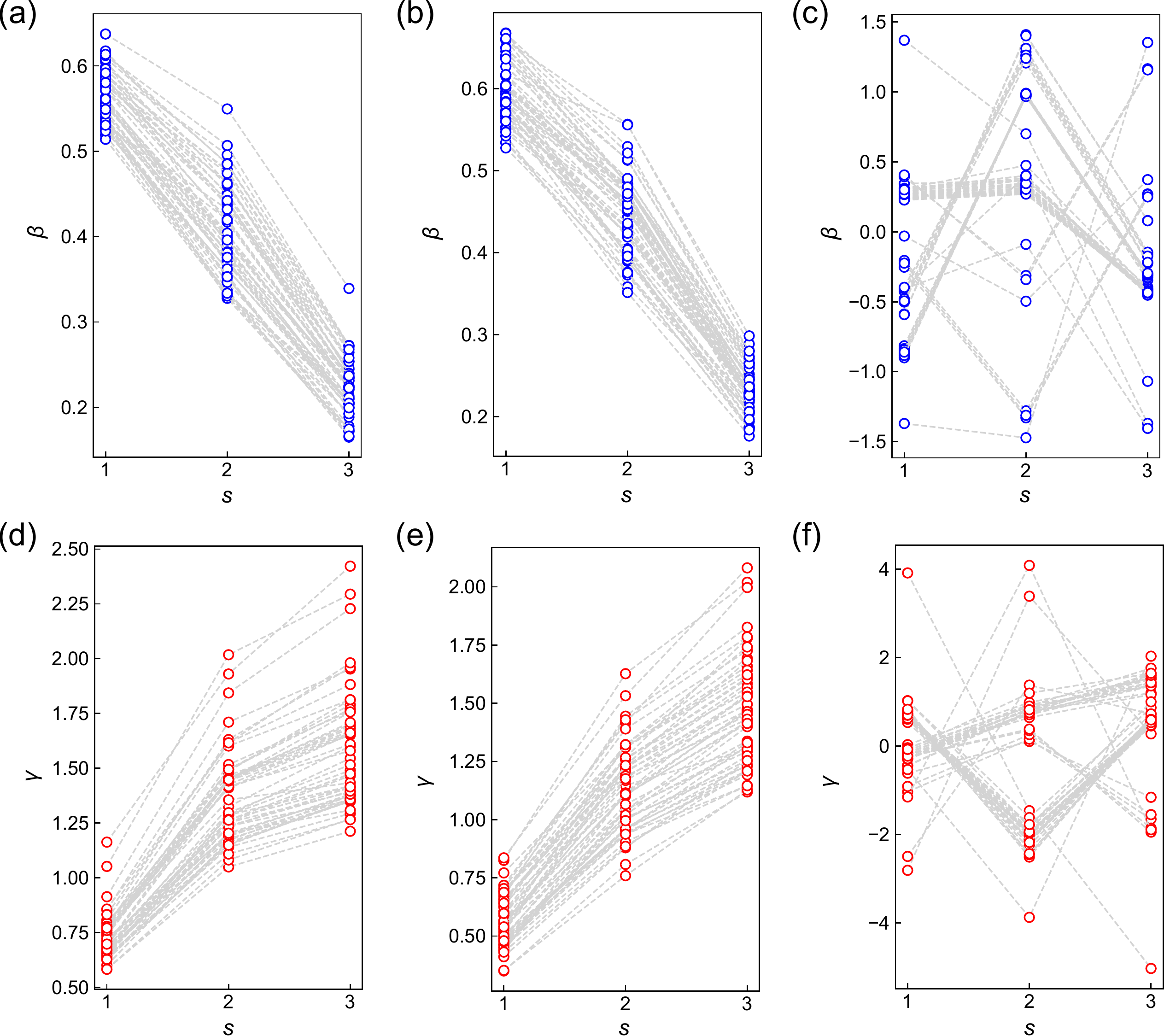}
\end{center}
\caption{Optimized $\beta_s$ and $\gamma_s$ of (a, d) QAOA and WS-QAOA with (b, e) $\alpha=0.4$ and (c, f) $\alpha=1$ for $d=1$ on 50 instances of w3R graph ($n=10$). The ansatz depth is $p=3$. The parameters are optimized by INTERP in (a, b, d, e) and RI in (c, f).}
\label{fig:beta_gamma}
\end{figure}

In WS-QAOA, we set approximate solutions $\{x_i^0\}$ by flipping $d$ bits randomly selected from $n$ bits in the solution $\{x_i^{\rm sol}\}$.
In other words, $d$ represents the Hamming distance of $\{x_i^0\}$ to $\{x^{\rm sol}_i\}$.
Figures~\ref{fig:beta_gamma} show the optimized parameters of QAOA $(\alpha=0)$, and WS-QAOA with $\alpha=0.4, 1$ for $d=1$ on 50 graph instances of $n=10$. 
Figs.~\ref{fig:beta_gamma} (a) and \ref{fig:beta_gamma}(d) show that in QAOA, $\beta_s$ ($\gamma_s$) decreases (increases) with $s$, which resembles the process of QA~\cite{Zhou2020}. 
We observe a similar trend in WS-QAOA with $\alpha=0.4$ in Figs.~\ref{fig:beta_gamma}(b) and \ref{fig:beta_gamma}(e).
Meanwhile, when $\alpha=1$, the parameters are not monotonic against $s$, which may reflect that the property of QA declines as $\alpha$ becomes larger. 
We also note that the parameters are optimized by RI in Figs.~\ref{fig:beta_gamma}(c) and \ref{fig:beta_gamma}(f).

\begin{figure}[htb]
\begin{center}
\includegraphics[width=\columnwidth]{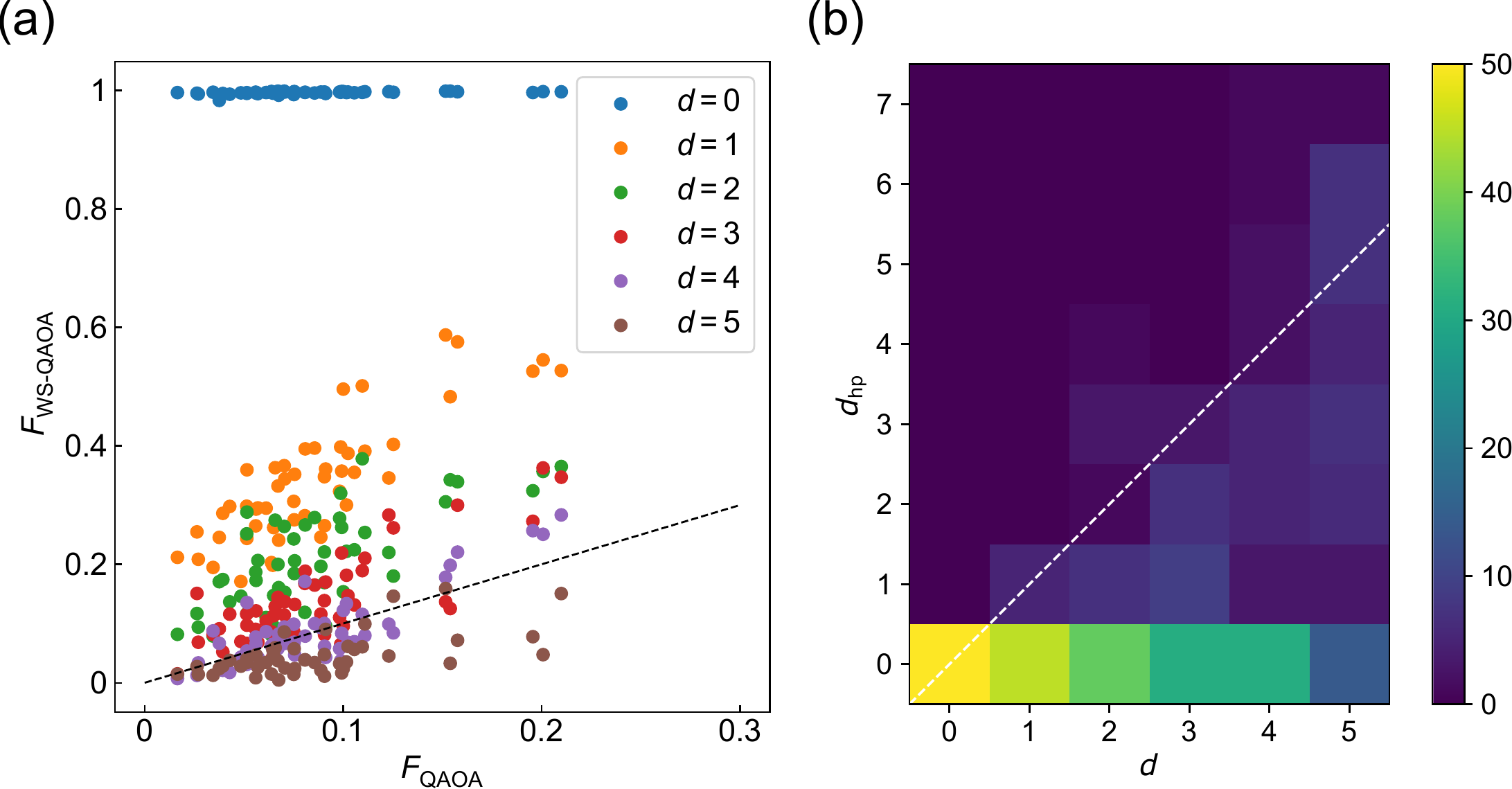}
\end{center}
\caption{Performance of WS-QAOA with $\alpha=0.4$ at $p=3$ on 50 instances of w3R graph ($n=14$). $d$ represents the Hamming distance between the approximate solution $\{x^0_i\}$ and exact one $\{x^{\rm sol}_i\}$. (a) Fidelity of WS-QAOA ($F_{\rm WS-QAOA}$) versus that of QAOA ($F_{\rm QAOA}$). The dotted line corresponds to $F_{\rm WS-QAOA}=F_{\rm QAOA}$. (b) Histogram over $d$ and $d_{\rm hp}$. $d_{\rm hp}$ is the Hamming distance of $\{x^{\rm hp}_i\}$ to $\{x^{\rm sol}_i\}$. The dotted line corresponds to $d_{\rm hp}=d$.}
\label{fig:F_vs_F_WS-QAOA}
\end{figure}

We compare the performance of WS-QAOA to that of QAOA.
As a performance indicator, we use the fidelity of the optimized ansatz $\ket{\Psi_{\rm WS-QAOA}}$. We define the fidelity of a wave function $\ket{\Phi}$ as 
\begin{equation}
F=|\braket{\{x^{\rm sol}_i\}|\Phi}|^2+|\braket{\{\overline{x}^{\rm sol}_i\}|\Phi}|^2.
\label{eq:fidelity}
\end{equation}
In Fig.~\ref{fig:F_vs_F_WS-QAOA}(a), $F$ of WS-QAOA with $\alpha=0.4$ at $p=3$ is plotted against that of QAOA on 50 graph instances of $n=14$. 
In the following, we focus on $\alpha=0.4, 1$. 
We refer to Appendix~A for a closer look at $\alpha$ dependence.
Figure~\ref{fig:F_vs_F_WS-QAOA}(a) indicates that the relative performance of WS-QAOA against QAOA is dominated by the Hamming distance $d$. 
Importantly, $F_{\rm WS-QAOA}$ becomes higher as $d$ decreases.
We find that WS-QAOA outperforms QAOA in all cases for $d\leq2$ and in most cases for $d=3$ [Fig.~\ref{fig:F_vs_F_WS-QAOA}(a)]. 
We note that $F_{\rm WS-QAOA}$ is always almost unity for $d=0$.
The enhanced fidelity with the decrease in $d$ has also been observed in BQA~\cite{Grass2019}.

Success of WS-QAOA with small $d$ is also manifested in the bit string with highest probability, $\{x^{\rm hp}_i\}$, in the optimized ansatz.
Figure~\ref{fig:F_vs_F_WS-QAOA}(b) shows the histogram over $d$ and the Hamming distance of $\{x^{\rm hp}_i\}$ to $\{x^{\rm sol}_i\}$, $d_{\rm hp}$, on $50$ instances of $n=14$.
One can see that the closer the approximate solution is to the exact solution, the more likely $\{x^{\rm hp}_i\}$ is to correspond with the exact solution ($d_{\rm hp}=0$) [Fig.~\ref{fig:F_vs_F_WS-QAOA}(b)]. 
It is notable that for $d=4$, the optimized ansatz still yields the solution as the highest-probability string in about a half of the instances. 

\begin{figure}[htb]
\begin{center}
\includegraphics[width=\columnwidth]{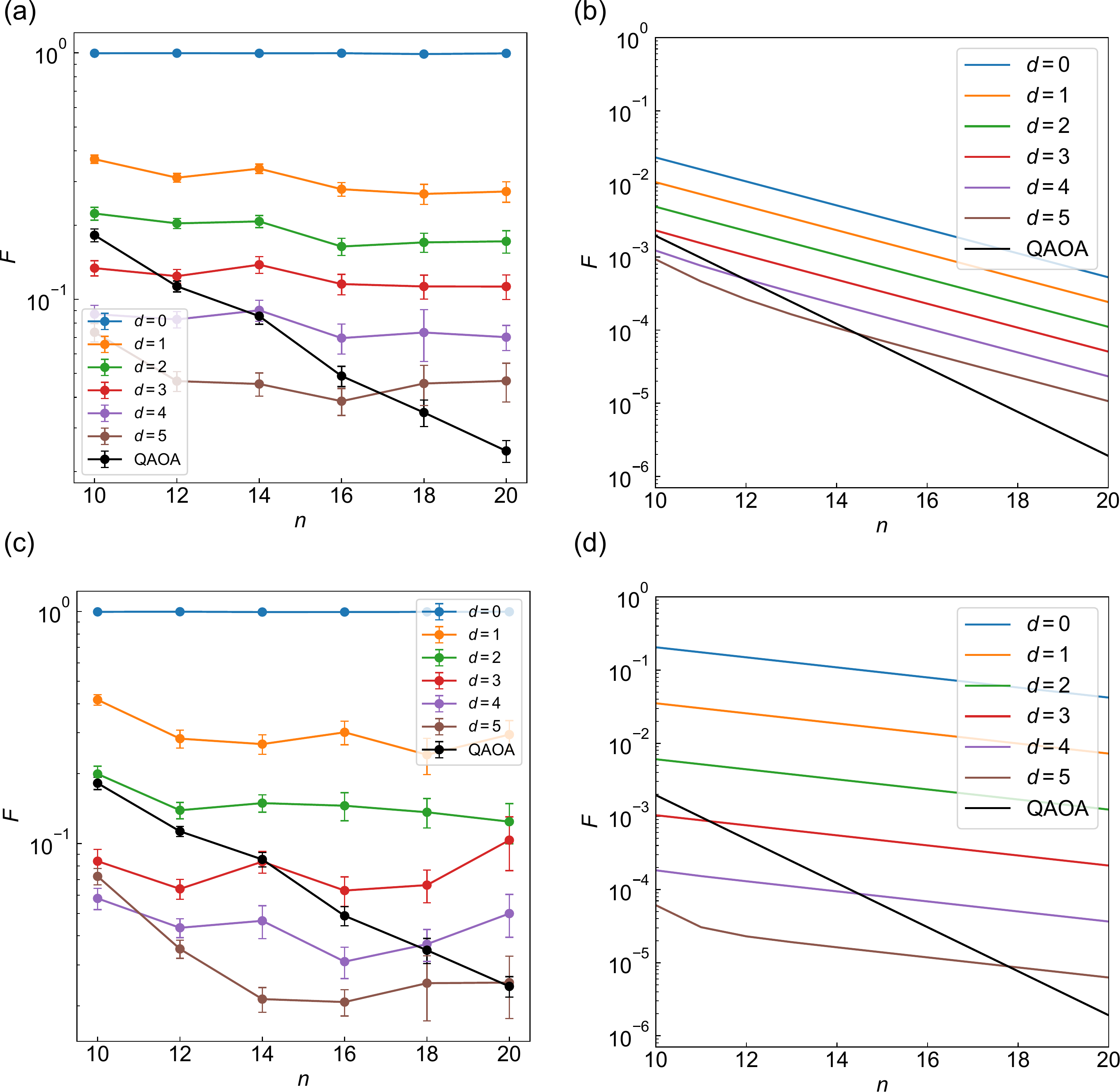}
\end{center}
\caption{(a, c) Graph size dependence of the average fidelity obtained by WS-QAOA with different $d$ compared to QAOA ($\alpha=0$) for $p=3$. (a) corresponds to $\alpha=0.4$ and (c) to $\alpha=1$. The fidelity is averaged over 50 instances for $n\leq 14$, 20 for $n=16$, 15 for $n=18$, and 10 for $n=20$. The error bar represents s.e.m. (b, d) Calculated fidelity for the initial state of the ansatz, Eq.~(\ref{eq:F_initial_state}).}
\label{fig:F_vs_n_WS-QAOA}
\end{figure}

We proceed to study the graph size dependence. 
In Figs.~\ref{fig:F_vs_n_WS-QAOA}(a) and \ref{fig:F_vs_n_WS-QAOA}(c), the averaged fidelity of WS-QAOA at $p=3$ is shown against the number of vertices $n$, together with that of QAOA ($\alpha=0$).
Figures~\ref{fig:F_vs_n_WS-QAOA}(a) and \ref{fig:F_vs_n_WS-QAOA}(c) correspond to $\alpha=0.4$ and $\alpha=1$, respectively.
We present the entire data of $p=1–4$ in Appendix~B.
In both WS-QAOA and QAOA, $F$ shows a nearly exponential decay with $n$, but importantly it decreases less steeply in WS-QAOA than in QAOA.
As a result, with larger $n$, WS-QAOA outperforms QAOA with even larger $d$.  
We also compare $\alpha=0.4$ and $\alpha=1$.
Figures~\ref{fig:F_vs_n_WS-QAOA}(a) and \ref{fig:F_vs_n_WS-QAOA}(c) indicate that as $d$ increases incrementally, fidelity decreases roughly by a constant multiplicative factor (aside from $d=0\to 1$) and that the factor is smaller for $\alpha=0.4$ than for $\alpha=1$. 
These features seem to stem from the initial state at least in part.
In Figs.~\ref{fig:F_vs_n_WS-QAOA}(b) and \ref{fig:F_vs_n_WS-QAOA}(d), we present the fidelity of $\ket{\Psi_0}$ for $\alpha=0.4$ and $\alpha=1$, which is derived from Eqs.~(\ref{eq:fidelity}) and (\ref{eq:WS-QAOA_initial_state}) with $\ket{\Phi}=\ket{\Psi_0}$ as 
\begin{equation}
\begin{split}
F_0&=\cos^{2d}\left(\frac{\pi}{4}+\frac{\tan^{-1}\alpha}{2}\right)\cos^{2(n-d)}\left(\frac{\pi}{4}-\frac{\tan^{-1}\alpha}{2}\right)\\
&+\cos^{2d}\left(\frac{\pi}{4}-\frac{\tan^{-1}\alpha}{2}\right)\cos^{2(n-d)}\left(\frac{\pi}{4}+\frac{\tan^{-1}\alpha}{2}\right).
\end{split}
\label{eq:F_initial_state}
\end{equation}
In Figs.~\ref{fig:F_vs_n_WS-QAOA}(b) and \ref{fig:F_vs_n_WS-QAOA}(d), one can observe similar behaviors to Figs.~\ref{fig:F_vs_n_WS-QAOA}(a) and \ref{fig:F_vs_n_WS-QAOA}(c), although the magnitude of the fidelity is significantly improved by the optimized circuit. 

\begin{figure}[htb]
\begin{center}
\includegraphics[width=\columnwidth]{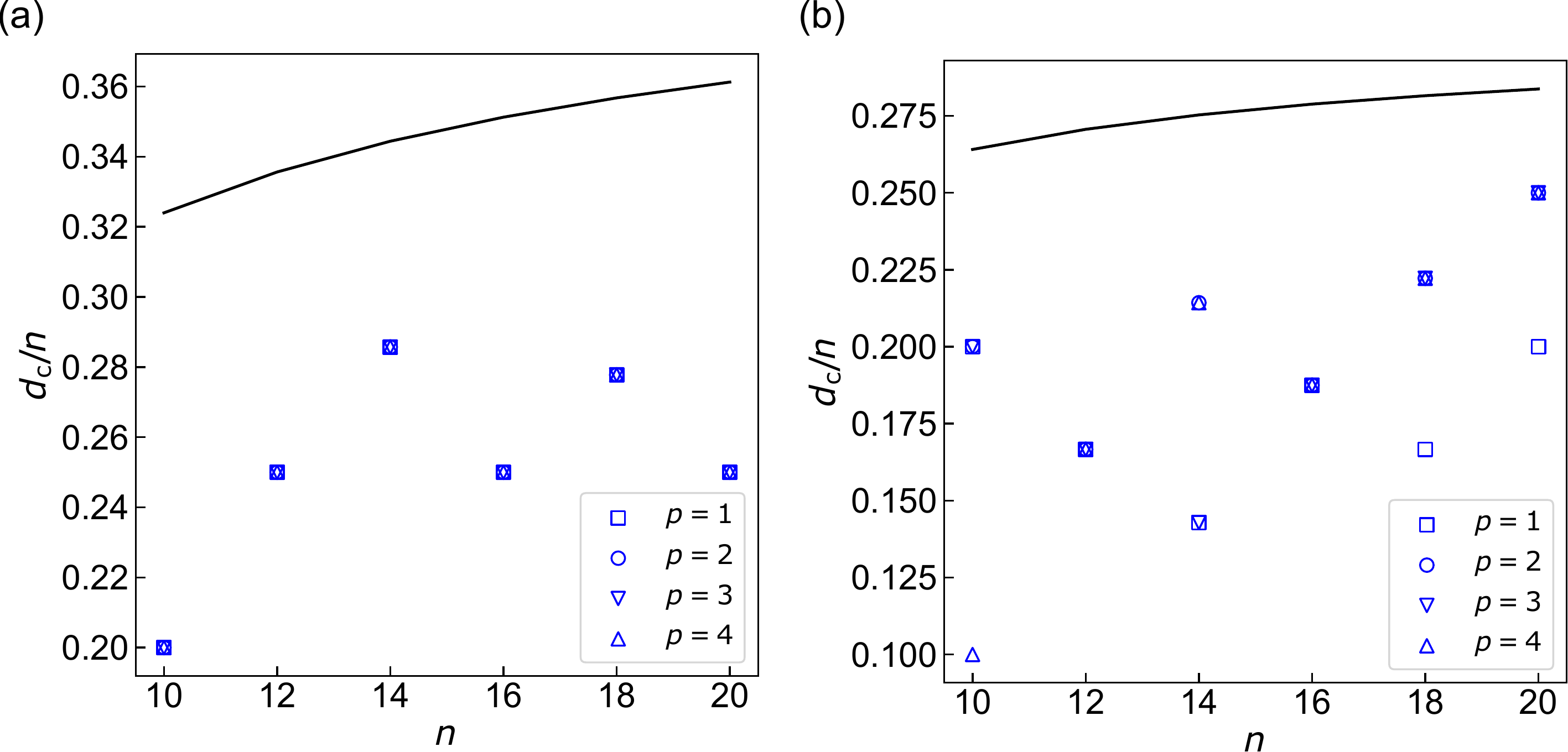}
\end{center}
\caption{Critical relative Hamming distance $d_{\rm c}/n$ plotted against $n$ for (a) $\alpha=0.4$ and (b) $\alpha=1$. When approximate solutions are less-than-$d_{\rm c}$ away from exact solutions, WS-QAOA outperforms QAOA on average. The black line denotes the theoretical value of $d_{\rm c}/n$ estimated for the initial state $\ket{\Psi_0}$ (see Eq.~(\ref{eq:dc_over_n_plus}) in the text).}
\label{fig:dc_over_n}
\end{figure}

The calculations above indicate how close approximate solutions should be to the exact solutions for WS-QAOA to outperform QAOA.
From the graph size dependence of the fidelity [Figs.~\ref{fig:F_vs_n_WS-QAOA}(a) and \ref{fig:F_vs_n_WS-QAOA}(c)], we estimate the critical Hamming distance of $\{x^0_i\}$, $d_{\rm c}$, which determines whether WS-QAOA outperforms QAOA or not.
For example, we estimate $d_{\rm c}=3$ for $n=12, \alpha=0.4$ from Fig.~\ref{fig:F_vs_n_WS-QAOA}(a).
Figures~\ref{fig:dc_over_n}(a) and \ref{fig:dc_over_n}(b) show $d_{\rm c}$ scaled by the graph size $n$ for $\alpha=0.4$ and $\alpha=1$, respectively.
For $\alpha=0.4$, $d_{\rm c}/n$ ranges within $[0.2, 0.3)$, whereas for $\alpha=1$, it hovers from $0.1$ to $0.25$.
We also theoretically derive $d_{\rm c}/n$ for the initial state $\ket{\Psi_0}$ starting from $F_0(\alpha)=F_0(\alpha=0)$ (see Appendix C), which reads
\begin{equation}
\begin{split}
\frac{d_{\rm c}}{n}&=\frac{\log\left(\cos\frac{\pi}{4}/\cos\left(\frac{\pi}{4}-\frac{\tan^{-1}\alpha}{2}\right)\right)}{\log\tan\left(\frac{\pi}{4}-\frac{\tan^{-1}\alpha}{2}\right)}\\
&+\frac{1}{2n}\frac{\log\left(1+\sqrt{1-\sin^{2n}\left(\frac{\pi}{2}-\tan^{-1}\alpha\right)}\right)}{\log\tan\left(\frac{\pi}{4}-\frac{\tan^{-1}\alpha}{2}\right)}.
\end{split}
\label{eq:dc_over_n_plus}
\end{equation}
We draw the curve of Eq.~(\ref{eq:dc_over_n_plus}) in Figs.~\ref{fig:dc_over_n}. 
One can see that in both $\alpha$, $d_{\rm c}/n$ estimated from the actual fidelity is smaller than the theoretical curve for the initial state.
This indicates that QAOA gains more fidelity by the optimized unitary circuit than WS-QAOA. 

\section{WS-QAOA combined with QAOA}\label{sec:QAOA+WS}

\begin{figure}[htb]
\begin{center}
\includegraphics[width=\columnwidth]{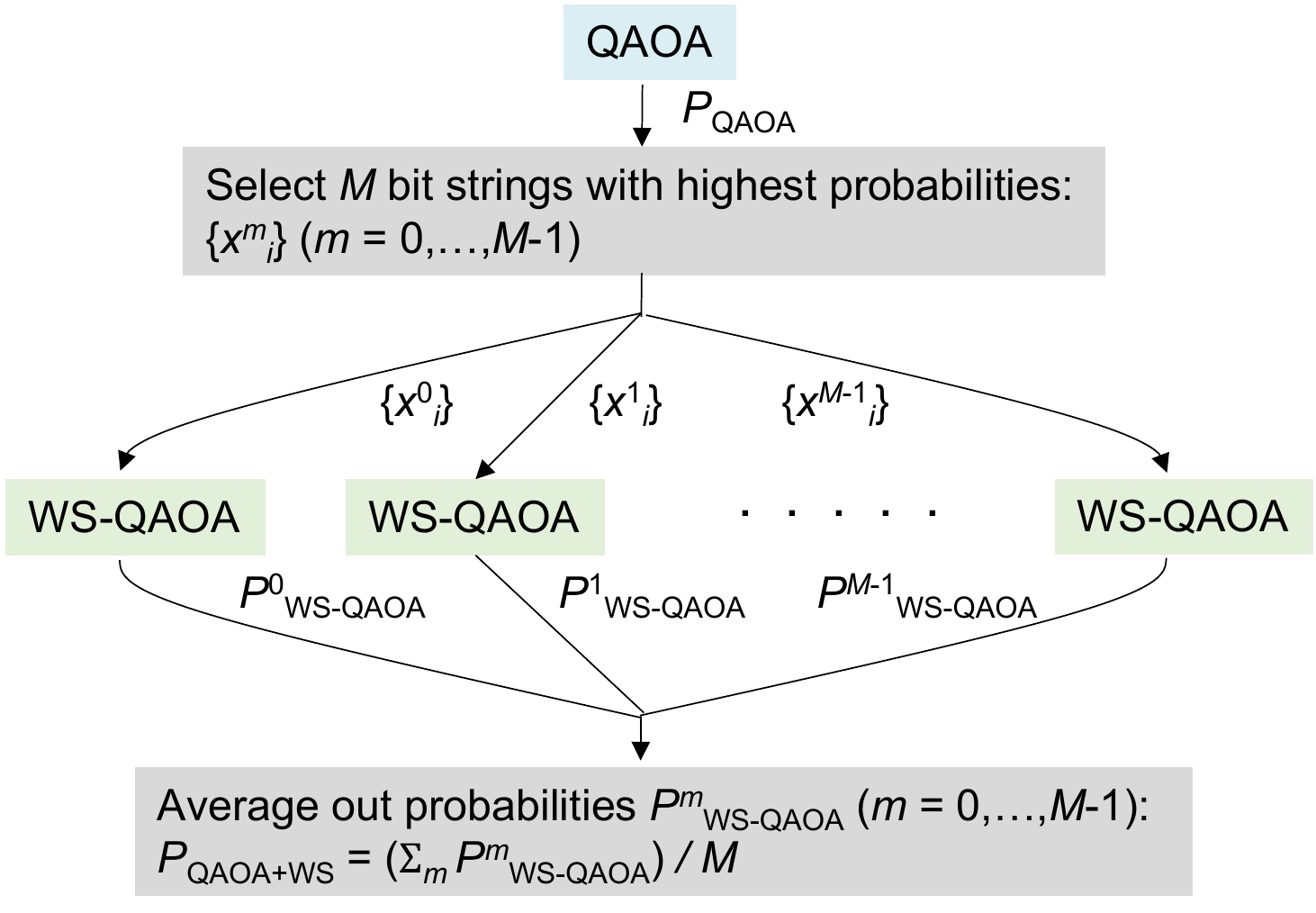}
\end{center}
\caption{Flow diagram of QAOA+WS.}
\label{fig:QAOA+WS_schematic}
\end{figure}

In the previous section, we studied the dependence of the WS-QAOA performance on approximate solutions and revealed that their Hamming distance to the exact solutions plays a crucial role.
In this section, we solve the MAX-CUT problem with WS-QAOA while finding suitable approximate solutions.  
To find them, as in the previous studies~\cite{Egger2021, Tate2020}, one could rely on the well-known classical algorithm for combinatorial optimization~\cite{Goemans1995}.
Instead, we employ QAOA here.
This resembles the approach in the previous study of BQA, where approximate solutions are obtained by QA beforehand~\cite{Grass2019}.

In Fig.~\ref{fig:QAOA+WS_schematic}, we depict a flow diagram of our procedure.
We call this procedure QAOA+WS hereafter.
First, we solve the problem using QAOA and pick up $M$ bit strings with highest probabilities, $\{x^m_i\}$ ($m=0,...,M-1$), based on the distribution $P_{\rm QAOA}$ from $\ket{\Psi_{\rm QAOA}}$.
Then we conduct WS-QAOA with $\{x^m_i\}$ as an approximate solution and obtain the distribution $P^m_{\rm WS-QAOA}$ from $\ket{\Psi_{\rm WS-QAOA}}$.
At the end, we obtain the final distribution $P_{\rm QAOA+WS}$ by averaging out $M$ distributions $P^m_{\rm WS-QAOA}$.

\begin{figure}[htb]
\begin{center}
\includegraphics[width=\columnwidth]{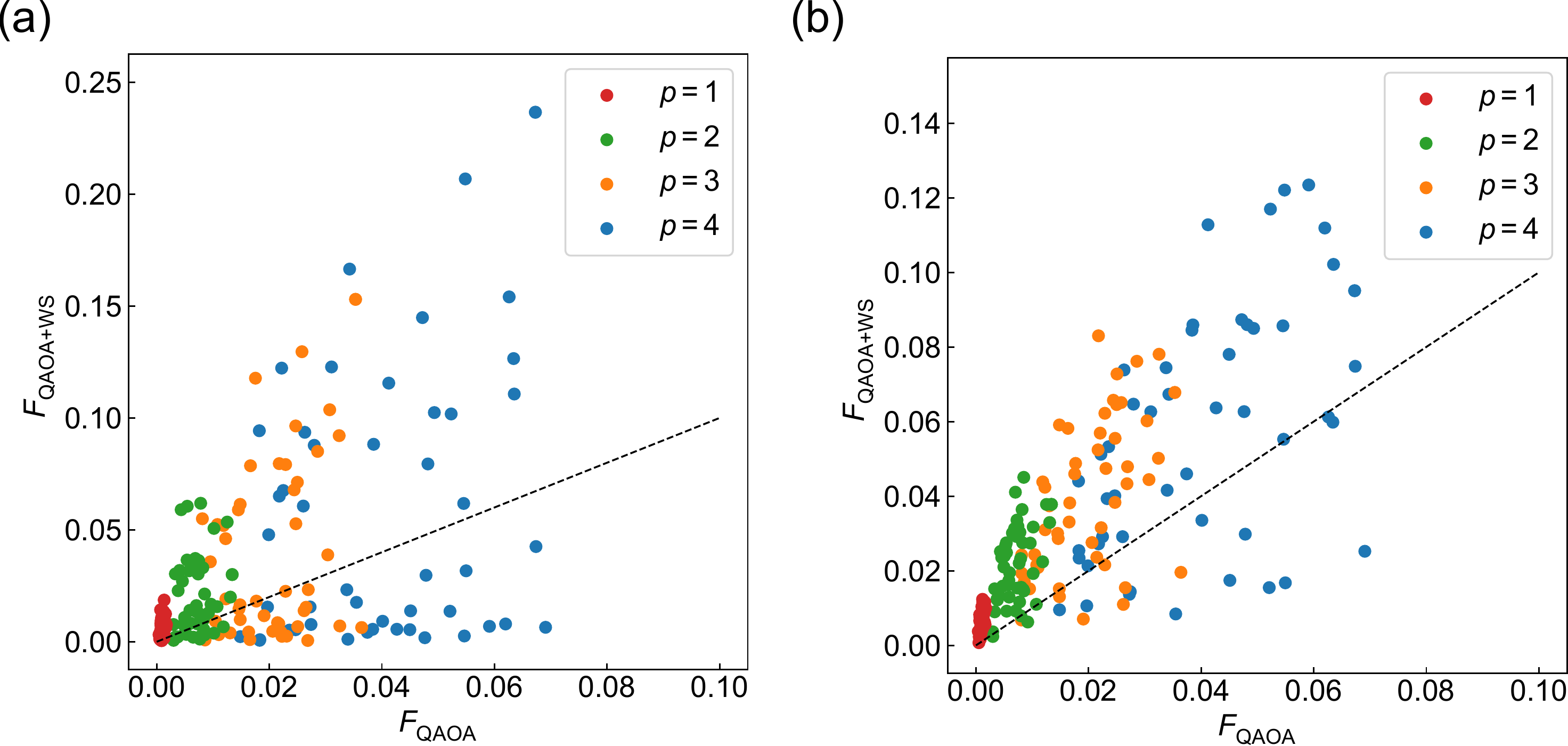}
\end{center}
\caption{Fidelity of QAOA+WS versus that of QAOA for 50 instances of $n=20$ with (a) $M=3$ and (b) $M=8$. For WS-QAOA in QAOA+WS, we set $\alpha=0.4$.}
\label{fig:F_vs_F_QAOA+WS}
\end{figure}

We compare the fidelity of QAOA+WS to that of QAOA.
The fidelity of QAOA+WS is calculated as the average over the fidelities of $M$ runs of WS-QAOA. 
Figures~\ref{fig:F_vs_F_QAOA+WS} present fidelities of QAOA+WS with $\alpha=0.4$ against those of QAOA over 50 graph instances of $n=20$. 
In Figs.~\ref{fig:F_vs_F_QAOA+WS}~(a) and \ref{fig:F_vs_F_QAOA+WS}(b), we set $M=3$ and $M=8$ for QAOA+WS, respectively. 
We note that if the exact solutions are included in $M$ approximate solutions, we drop them off, considering that $d=0$ almost always yields the perfect fidelity [Figs.~\ref{fig:F_vs_F_WS-QAOA}(a), \ref{fig:F_vs_n_WS-QAOA}(a)]. 
For $M=3$, QAOA+WS shows a sizable variance of the fidelity especially as $p$ increases. It shows a better performance for most instances than QAOA at $p=1$, but not necessarily at $p\geq 2$.
Meanwhile, for $M=8$, QAOA+WS shows a smaller variance and outperforms QAOA at $p\leq 3$ for most instances.
Smaller variance with larger $M$ seems to be natural, because the approximate solutions are more likely to have a wide range of the Hamming distance to the exact solutions as $M$ increases. 

\begin{figure}[htb]
\begin{center}
\includegraphics[width=\columnwidth]{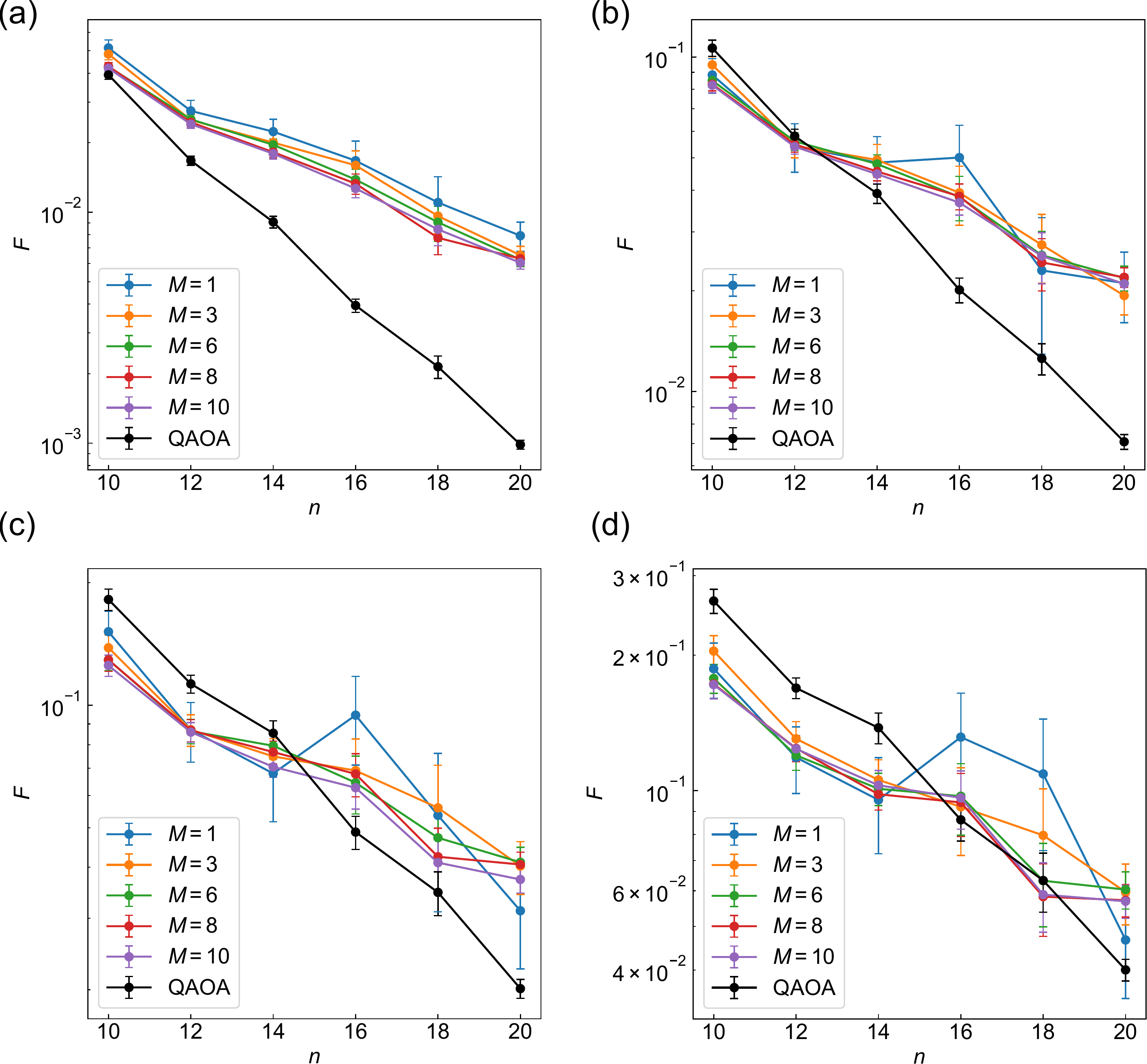}
\end{center}
\caption{Graph size dependence of the fidelities of QAOA+WS along with QAOA at (a–d) $p=1–4$. The fidelity is averaged over 50 graph instances for $n=10, 12, 14, 20$, 20 for $n=16$, and $15$ for $n=18$. For WS-QAOA in QAOA+WS, we set $\alpha=0.4$. The error bar represents s.e.m.}
\label{fig:F_vs_n_QAOA+WS}
\end{figure}

We also study the graph size dependence.
Figures~\ref{fig:F_vs_n_QAOA+WS} present the fidelity of QAOA+WS with $\alpha=0.4$ along with that of QAOA plotted against the number of vertices. 
In Figs.~\ref{fig:F_vs_n_QAOA+WS} (a–d), we set $p=1–4$, respectively.
The fidelity is averaged over 50 graph instances for $n=10, 12, 14, 20$, 20 for $n=16$, and 15 for $n=18$.
Importantly, the fidelity decays more slowly with $n$ in QAOA+WS than in QAOA for $p=1–4$ [Figs.~\ref{fig:F_vs_n_QAOA+WS}].
As a result, QAOA+WS on average outperforms QAOA for all $M$ as $n$ increases; for $n\geq 10$ at $p=1$, $n\geq 14$ at $p=2$, $n\geq 16$ at $p=3$, and $n=20$ at $p=4$.
It should be also mentioned that QAOA+WS becomes more beneficial for smaller $p$, because the difference in the decay with $n$ seems to decrease as $p$ increases.

\section{Conclusion}\label{sec:conclusion}
In this work, we systematically studied how the performance of WS-QAOA depends on the quality of approximate solutions by numerical simulations on the MAX-CUT problem on w3R graphs.
We found that WS-QAOA yields higher fidelities than QAOA when one uses approximate solutions that are close enough to the exact solutions in terms of the Hamming distance; WS-QAOA with $\alpha=0.4$ ($\alpha=1$) on average outperforms QAOA if the relative Hamming distance of approximate solutions to the exact ones, $d_{\rm c}/n$, is below $0.2–0.3$ ($0.1–0.25$).
We also obtained theoretical curves that explain those properties.
Lastly, we showed that QAOA could serve as a capable way to find approximate solutions for WS-QAOA.
We found out that WS-QAOA combined with QAOA shows a better performance than QAOA specifically when the depth is limited to a small number. 

We believe that our findings could allow one to make a clear understanding of the efficacy of WS-QAOA. They might also be helpful to determining the criteria of approximate solutions for WS-QAOA.

\begin{acknowledgments}
This work was supported by MEXT via the ``Program for Promoting Researches on the Supercomputer Fugaku'' (JP-MXP1020200205) and JSPS KAKENHI via the ``Grant-in-Aid for Scientific Research(A)'' Grant Number 21H04553.
This work was also supported in part by MEXT Quantum Leap Flagship Program (MEXTQLEAP) Grant No.~JPMXS0120319794.
Part of numerical calculation was carried out at the Supercomputer Center, Institute for Solid State Physics, University of Tokyo.
\end{acknowledgments}

\appendix*
\section*{Appendix A: $\alpha$ dependence of the fidelity of WS-QAOA}
\label{sec:fidelity_alpha_dep}

\begin{figure}[htb]
\begin{center}
\includegraphics[width=\columnwidth]{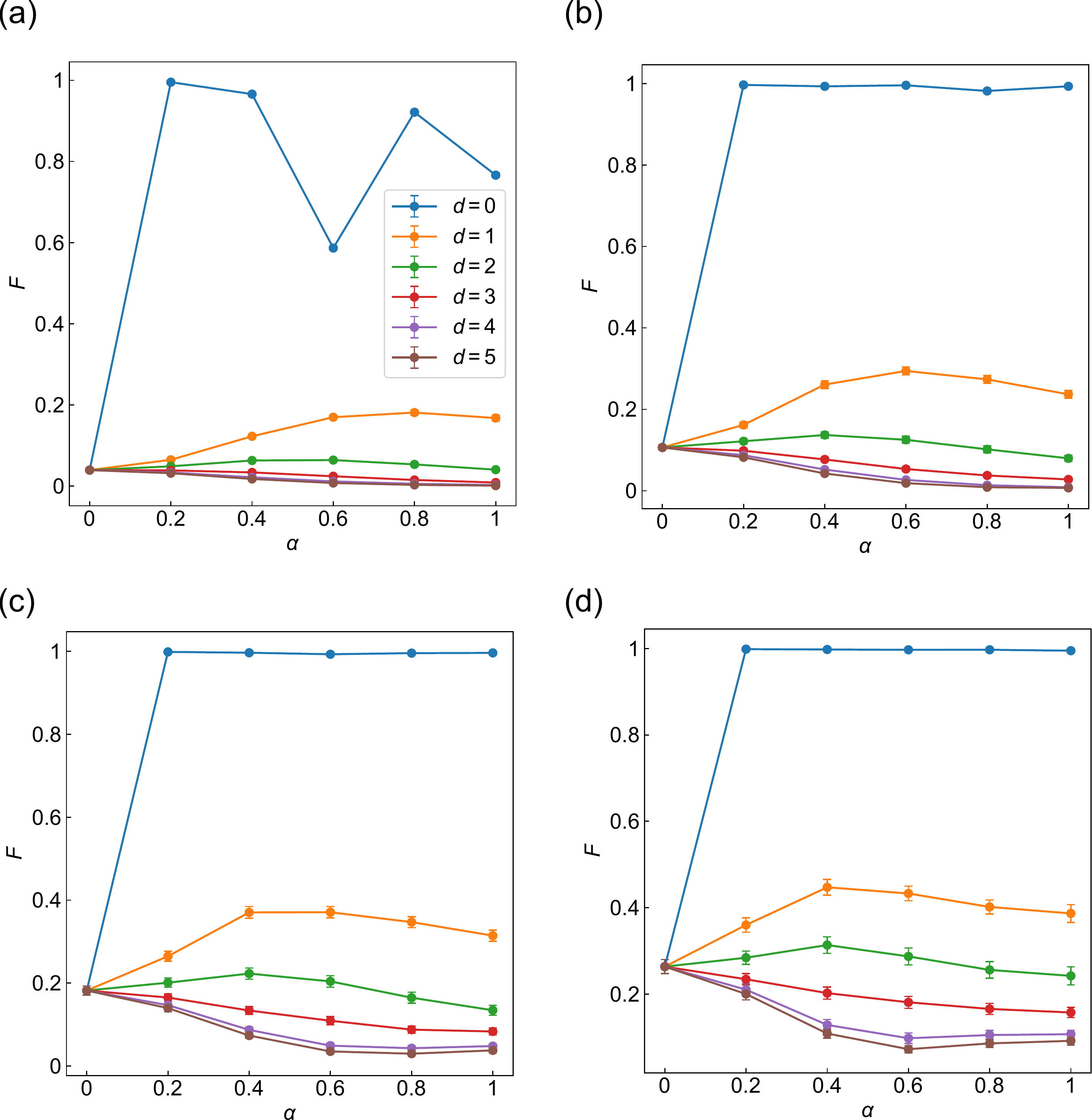}
\end{center}
\caption{$\alpha$ dependence of the fidelities for WS-QAOA of $n=10$ with (a) $p=1$, (b) $p=2$, (c) $p=3$, and (d) $p=4$. The fidelity is averaged over 50 graph instances. The error bar represents s.e.m. For optimization, we use RI for $d=0$ ($\alpha\neq0$) and INTERP for the other cases.}
\label{fig:F_vs_alpha_WS-QAOA}
\end{figure}

In this Appendix, we systematically study $\alpha$ dependence of WS-QAOA.
Figures~\ref{fig:F_vs_alpha_WS-QAOA}(a–d) display the averaged fidelity over 50 graph instances of $n=10$ as a function of $\alpha$ for $p=1–4$.
Here we optimize the parameters using RI for WS-QAOA with $d=0$ and INTERP otherwise.
We note that $\alpha=0$ corresponds to QAOA.
For WS-QAOA with $d=0$, the fidelity $F$ almost equals 1, aside from $p=1$.
For $d=1, 2$, $F$ has a peak around $\alpha=0.4–0.8$, whereas, for $d\geq3$, $F$ monotonically decreases with $\alpha$.
Therefore the optimal $\alpha$ varies with $d$.

\section*{Appendix B: Comparison of the fidelities of WS-QAOA with different $p$}
\label{sec:fidelity_all_p}
\begin{figure*}[htb]
\begin{center}
\includegraphics[width=\linewidth]{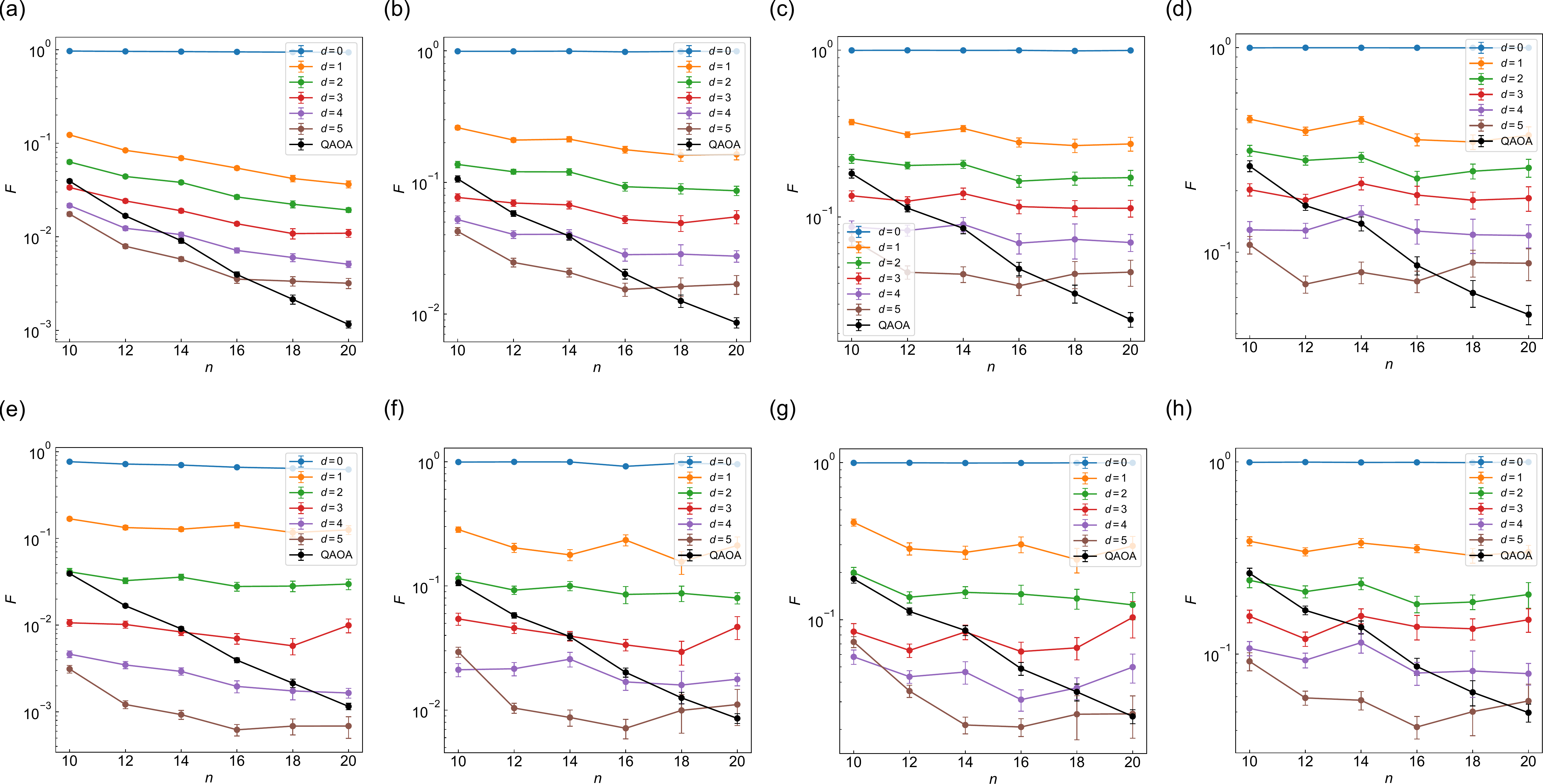}
\end{center}
\caption{Graph size dependence of the fidelities of WS-QAOA along with QAOA (a–d) $\alpha=0.4$ and (e–h) $\alpha=1$. (a) and (e) correspond to $p=1$, (b) and (f) to $p=2$, (c) and (g) to $p=3$, and (d) and (h) to $p=4$. The fidelity is averaged over 50 graph instances for $n\leq14$, 20 for $n=16$, 15 for $n=18$, and 10 for $n=20$. The error bar stands for s.e.m.}
\label{fig:F_vs_n_WS-QAOA_supple}
\end{figure*}

In this Appendix, we compare the fidelities of WS-QAOA with different values of $p$.
Figures~\ref{fig:F_vs_n_WS-QAOA_supple}(a–d) show the $n$ dependence of the fidelity with $p=1–4$ for $\alpha=0.4$.
Figures~\ref{fig:F_vs_n_WS-QAOA_supple}(e-h) correspond to $\alpha=1$. 
Note that Figs.~\ref{fig:F_vs_n_WS-QAOA_supple}(c) and \ref{fig:F_vs_n_WS-QAOA_supple}(g) are identical to Figs.~\ref{fig:F_vs_n_WS-QAOA}(a) and \ref{fig:F_vs_n_WS-QAOA}(c) in the main text.
As expected, Figures~\ref{fig:F_vs_n_WS-QAOA_supple} show that $F$ increases with $p$ for both $\alpha=0.4$ and $\alpha=1$.
One can see that the tendencies mentioned in Sec.~\ref{sec:WS-QAOA_numerics} are shared among all $p$. In other words, the fidelity of WS-QAOA decays more slowly with $n$ than that of QAOA and decreases by a multiplicative factor with $d$, which seems to be larger for $\alpha=0.4$ than $\alpha=1$. 

\section*{Appendix C: Derivation of $d_{\rm c}/n$ for the initial state of the WS-QAOA ansatz} 
\label{sec:derivation_dc_over_n}
Here we show derivation of Eq.~(\ref{eq:dc_over_n_plus}).
From Eq.~(\ref{eq:F_initial_state}), $F_0(\alpha)=F_0(\alpha=0)$ yields
\begin{equation}
\begin{split}
\cos^{2d}&\left(\frac{\pi}{4}+\frac{\theta}{2}\right)\cos^{2(n-d)}\left(\frac{\pi}{4}-\frac{\theta}{2}\right)+\\
&\cos^{2d}\left(\frac{\pi}{4}-\frac{\theta}{2}\right)\cos^{2(n-d)}\left(\frac{\pi}{4}+\frac{\theta}{2}\right)=2\cos^{2n}\left(\frac{\pi}{4}\right),
\end{split}
\label{eq:F_alpha=F_0}
\end{equation}
where we set $\theta=\tan^{-1}\alpha$.
With $\delta$ defined as $\delta=\tan\left(\frac{\pi}{4}-\frac{\theta}{2}\right)$, Eq.~(\ref{eq:F_alpha=F_0}) is represented as
\begin{equation}
\cos^{2n}\left(\frac{\pi}{4}-\frac{\theta}{2}\right)\delta^{2d}+\frac{\cos^{2n}\left(\frac{\pi}{4}+\frac{\theta}{2}\right)}{\delta^{2d}}=2\cos^{2n}\left(\frac{\pi}{4}\right).
\label{eq:F_alpha=F_0_transformed}
\end{equation}
We solve Eq.~(\ref{eq:F_alpha=F_0_transformed}) with respect to $\delta^{2d}$ and obtain
\begin{equation}
\delta^{2d}=\frac{\cos^{2n}\left(\frac{\pi}{4}\right)}{\cos^{2n}\left(\frac{\pi}{4}-\frac{\theta}{2}\right)}\left(1\pm\sqrt{1-\sin^{2n}\left(\frac{\pi}{2}-\theta\right)}\right).
\label{eq:delta^2d}
\end{equation}
Logarithm of Eq.~(\ref{eq:delta^2d}) gives 
\begin{equation}
\begin{split}
2d\log\delta&=2n\log\frac{\cos\left(\frac{\pi}{4}\right)}{\cos\left(\frac{\pi}{4}-\frac{\theta}{2}\right)}\\
&+\log\left(1\pm\sqrt{1-\sin^{2n}\left(\frac{\pi}{2}-\theta\right)}\right).
\end{split}
\end{equation}
Thus $d/n$ is derived as
\begin{equation}
\begin{split}
\frac{d}{n}&=\frac{\log\left(\cos\frac{\pi}{4}/\cos\left(\frac{\pi}{4}-\frac{\theta}{2}\right)\right)}{\log\delta}\\
&+\frac{1}{2n}\frac{\log\left(1\pm\sqrt{1-\sin^{2n}\left(\frac{\pi}{2}-\theta\right)}\right)}{\log\delta}.
\end{split}
\label{eq:d_over_n}
\end{equation}
As we denote the lefthand side of Eq.~(\ref{eq:d_over_n}) as $d^\pm_{\rm c}/n$ corresponding to $\pm$ in the righthand side, $d^+_{\rm c}/n+d^-_{\rm c}/n=1$ holds.
Considering $0\leq d\leq n/2$ without loss of generality, $F_0(\alpha)\geq F_0(\alpha=0)$ holds for $d\leq d^+_{\rm c}$.
Hence we represent $d^+_{\rm c}/n$ in Eq.~(\ref{eq:dc_over_n_plus}).

\bibliography{citation} 

\begin{thebibliography}{26}%
\makeatletter
\providecommand \@ifxundefined [1]{%
 \@ifx{#1\undefined}
}%
\providecommand \@ifnum [1]{%
 \ifnum #1\expandafter \@firstoftwo
 \else \expandafter \@secondoftwo
 \fi
}%
\providecommand \@ifx [1]{%
 \ifx #1\expandafter \@firstoftwo
 \else \expandafter \@secondoftwo
 \fi
}%
\providecommand \natexlab [1]{#1}%
\providecommand \enquote  [1]{``#1''}%
\providecommand \bibnamefont  [1]{#1}%
\providecommand \bibfnamefont [1]{#1}%
\providecommand \citenamefont [1]{#1}%
\providecommand \href@noop [0]{\@secondoftwo}%
\providecommand \href [0]{\begingroup \@sanitize@url \@href}%
\providecommand \@href[1]{\@@startlink{#1}\@@href}%
\providecommand \@@href[1]{\endgroup#1\@@endlink}%
\providecommand \@sanitize@url [0]{\catcode `\\12\catcode `\$12\catcode
  `\&12\catcode `\#12\catcode `\^12\catcode `\_12\catcode `\%12\relax}%
\providecommand \@@startlink[1]{}%
\providecommand \@@endlink[0]{}%
\providecommand \url  [0]{\begingroup\@sanitize@url \@url }%
\providecommand \@url [1]{\endgroup\@href {#1}{\urlprefix }}%
\providecommand \urlprefix  [0]{URL }%
\providecommand \Eprint [0]{\href }%
\providecommand \doibase [0]{http://dx.doi.org/}%
\providecommand \selectlanguage [0]{\@gobble}%
\providecommand \bibinfo  [0]{\@secondoftwo}%
\providecommand \bibfield  [0]{\@secondoftwo}%
\providecommand \translation [1]{[#1]}%
\providecommand \BibitemOpen [0]{}%
\providecommand \bibitemStop [0]{}%
\providecommand \bibitemNoStop [0]{.\EOS\space}%
\providecommand \EOS [0]{\spacefactor3000\relax}%
\providecommand \BibitemShut  [1]{\csname bibitem#1\endcsname}%
\let\auto@bib@innerbib\@empty
\bibitem [{\citenamefont {Kjaergaard}\ \emph {et~al.}(2020)\citenamefont
  {Kjaergaard}, \citenamefont {Schwartz}, \citenamefont {Braumüller},
  \citenamefont {Krantz}, \citenamefont {Wang}, \citenamefont {Gustavsson},\
  and\ \citenamefont {Oliver}}]{Kjaergaard2020}%
  \BibitemOpen
  \bibfield  {author} {\bibinfo {author} {\bibfnamefont {M.}~\bibnamefont
  {Kjaergaard}}, \bibinfo {author} {\bibfnamefont {M.~E.}\ \bibnamefont
  {Schwartz}}, \bibinfo {author} {\bibfnamefont {J.}~\bibnamefont
  {Braumüller}}, \bibinfo {author} {\bibfnamefont {P.}~\bibnamefont {Krantz}},
  \bibinfo {author} {\bibfnamefont {J.~I.~J.}\ \bibnamefont {Wang}}, \bibinfo
  {author} {\bibfnamefont {S.}~\bibnamefont {Gustavsson}}, \ and\ \bibinfo
  {author} {\bibfnamefont {W.~D.}\ \bibnamefont {Oliver}},\ }\href@noop {}
  {\bibfield  {journal} {\bibinfo  {journal} {Annu. Rev. Condens. Matter
  Phys.}\ }\textbf {\bibinfo {volume} {11}},\ \bibinfo {pages} {369} (\bibinfo
  {year} {2020})}\BibitemShut {NoStop}%
\bibitem [{\citenamefont {Preskill}(2018)}]{Preskill2018}%
  \BibitemOpen
  \bibfield  {author} {\bibinfo {author} {\bibfnamefont {J.}~\bibnamefont
  {Preskill}},\ }\href@noop {} {\bibfield  {journal} {\bibinfo  {journal}
  {Quantum}\ }\textbf {\bibinfo {volume} {2}},\ \bibinfo {pages} {79} (\bibinfo
  {year} {2018})}\BibitemShut {NoStop}%
\bibitem [{\citenamefont {McArdle}\ \emph {et~al.}(2020)\citenamefont
  {McArdle}, \citenamefont {Endo}, \citenamefont {Aspuru-Guzik}, \citenamefont
  {Benjamin},\ and\ \citenamefont {Yuan}}]{McArdle2020}%
  \BibitemOpen
  \bibfield  {author} {\bibinfo {author} {\bibfnamefont {S.}~\bibnamefont
  {McArdle}}, \bibinfo {author} {\bibfnamefont {S.}~\bibnamefont {Endo}},
  \bibinfo {author} {\bibfnamefont {A.}~\bibnamefont {Aspuru-Guzik}}, \bibinfo
  {author} {\bibfnamefont {S.~C.}\ \bibnamefont {Benjamin}}, \ and\ \bibinfo
  {author} {\bibfnamefont {X.}~\bibnamefont {Yuan}},\ }\href@noop {} {\bibfield
   {journal} {\bibinfo  {journal} {Rev. Mod. Phys.}\ }\textbf {\bibinfo
  {volume} {92}},\ \bibinfo {pages} {015003} (\bibinfo {year}
  {2020})}\BibitemShut {NoStop}%
\bibitem [{\citenamefont {Cerezo}\ \emph {et~al.}(2021)\citenamefont {Cerezo},
  \citenamefont {Arrasmith}, \citenamefont {Babbush}, \citenamefont {Benjamin},
  \citenamefont {Endo}, \citenamefont {Fujii}, \citenamefont {McClean},
  \citenamefont {Mitarai}, \citenamefont {Yuan}, \citenamefont {Cincio},\ and\
  \citenamefont {Coles}}]{Cerezo2021}%
  \BibitemOpen
  \bibfield  {author} {\bibinfo {author} {\bibfnamefont {M.}~\bibnamefont
  {Cerezo}}, \bibinfo {author} {\bibfnamefont {A.}~\bibnamefont {Arrasmith}},
  \bibinfo {author} {\bibfnamefont {R.}~\bibnamefont {Babbush}}, \bibinfo
  {author} {\bibfnamefont {S.~C.}\ \bibnamefont {Benjamin}}, \bibinfo {author}
  {\bibfnamefont {S.}~\bibnamefont {Endo}}, \bibinfo {author} {\bibfnamefont
  {K.}~\bibnamefont {Fujii}}, \bibinfo {author} {\bibfnamefont {J.~R.}\
  \bibnamefont {McClean}}, \bibinfo {author} {\bibfnamefont {K.}~\bibnamefont
  {Mitarai}}, \bibinfo {author} {\bibfnamefont {X.}~\bibnamefont {Yuan}},
  \bibinfo {author} {\bibfnamefont {L.}~\bibnamefont {Cincio}}, \ and\ \bibinfo
  {author} {\bibfnamefont {P.~J.}\ \bibnamefont {Coles}},\ }\href@noop {}
  {\bibfield  {journal} {\bibinfo  {journal} {Nat. Rev. Phys.}\ }\textbf
  {\bibinfo {volume} {3}},\ \bibinfo {pages} {625} (\bibinfo {year}
  {2021})}\BibitemShut {NoStop}%
\bibitem [{\citenamefont {Bharti}\ \emph {et~al.}(2022)\citenamefont {Bharti},
  \citenamefont {Cervera-Lierta}, \citenamefont {Kyaw}, \citenamefont {Haug},
  \citenamefont {Alperin-Lea}, \citenamefont {Anand}, \citenamefont {Degroote},
  \citenamefont {Heimonen}, \citenamefont {Kottmann}, \citenamefont {Menke},
  \citenamefont {Mok}, \citenamefont {Sim}, \citenamefont {Kwek},\ and\
  \citenamefont {Aspuru-Guzik}}]{bharti2021}%
  \BibitemOpen
  \bibfield  {author} {\bibinfo {author} {\bibfnamefont {K.}~\bibnamefont
  {Bharti}}, \bibinfo {author} {\bibfnamefont {A.}~\bibnamefont
  {Cervera-Lierta}}, \bibinfo {author} {\bibfnamefont {T.~H.}\ \bibnamefont
  {Kyaw}}, \bibinfo {author} {\bibfnamefont {T.}~\bibnamefont {Haug}}, \bibinfo
  {author} {\bibfnamefont {S.}~\bibnamefont {Alperin-Lea}}, \bibinfo {author}
  {\bibfnamefont {A.}~\bibnamefont {Anand}}, \bibinfo {author} {\bibfnamefont
  {M.}~\bibnamefont {Degroote}}, \bibinfo {author} {\bibfnamefont
  {H.}~\bibnamefont {Heimonen}}, \bibinfo {author} {\bibfnamefont {J.~S.}\
  \bibnamefont {Kottmann}}, \bibinfo {author} {\bibfnamefont {T.}~\bibnamefont
  {Menke}}, \bibinfo {author} {\bibfnamefont {W.-K.}\ \bibnamefont {Mok}},
  \bibinfo {author} {\bibfnamefont {S.}~\bibnamefont {Sim}}, \bibinfo {author}
  {\bibfnamefont {L.-C.}\ \bibnamefont {Kwek}}, \ and\ \bibinfo {author}
  {\bibfnamefont {A.}~\bibnamefont {Aspuru-Guzik}},\ }\href@noop {} {\bibfield
  {journal} {\bibinfo  {journal} {Rev. Mod. Phys.}\ }\textbf {\bibinfo {volume}
  {94}},\ \bibinfo {pages} {015004} (\bibinfo {year} {2022})}\BibitemShut
  {NoStop}%
\bibitem [{\citenamefont {Peruzzo}\ \emph {et~al.}(2014)\citenamefont
  {Peruzzo}, \citenamefont {McClean}, \citenamefont {Shadbolt}, \citenamefont
  {Yung}, \citenamefont {Zhou}, \citenamefont {Love}, \citenamefont
  {Aspuru-Guzik},\ and\ \citenamefont {O’Brien}}]{Peruzzo2014}%
  \BibitemOpen
  \bibfield  {author} {\bibinfo {author} {\bibfnamefont {A.}~\bibnamefont
  {Peruzzo}}, \bibinfo {author} {\bibfnamefont {J.}~\bibnamefont {McClean}},
  \bibinfo {author} {\bibfnamefont {P.}~\bibnamefont {Shadbolt}}, \bibinfo
  {author} {\bibfnamefont {M.-H.}\ \bibnamefont {Yung}}, \bibinfo {author}
  {\bibfnamefont {X.-Q.}\ \bibnamefont {Zhou}}, \bibinfo {author}
  {\bibfnamefont {P.~J.}\ \bibnamefont {Love}}, \bibinfo {author}
  {\bibfnamefont {A.}~\bibnamefont {Aspuru-Guzik}}, \ and\ \bibinfo {author}
  {\bibfnamefont {J.~L.}\ \bibnamefont {O’Brien}},\ }\href@noop {} {\bibfield
   {journal} {\bibinfo  {journal} {Nat. Commun.}\ }\textbf {\bibinfo {volume}
  {5}},\ \bibinfo {pages} {4213} (\bibinfo {year} {2014})}\BibitemShut
  {NoStop}%
\bibitem [{\citenamefont {McClean}\ \emph {et~al.}(2016)\citenamefont
  {McClean}, \citenamefont {Romero}, \citenamefont {Babbush},\ and\
  \citenamefont {Aspuru-Guzik}}]{McClean2016}%
  \BibitemOpen
  \bibfield  {author} {\bibinfo {author} {\bibfnamefont {J.~R.}\ \bibnamefont
  {McClean}}, \bibinfo {author} {\bibfnamefont {J.}~\bibnamefont {Romero}},
  \bibinfo {author} {\bibfnamefont {R.}~\bibnamefont {Babbush}}, \ and\
  \bibinfo {author} {\bibfnamefont {A.}~\bibnamefont {Aspuru-Guzik}},\
  }\href@noop {} {\bibfield  {journal} {\bibinfo  {journal} {New J. Phys.}\
  }\textbf {\bibinfo {volume} {18}},\ \bibinfo {pages} {023023} (\bibinfo
  {year} {2016})}\BibitemShut {NoStop}%
\bibitem [{\citenamefont {Farhi}\ \emph {et~al.}(2014)\citenamefont {Farhi},
  \citenamefont {Goldstone},\ and\ \citenamefont {Gutmann}}]{Farhi2014}%
  \BibitemOpen
  \bibfield  {author} {\bibinfo {author} {\bibfnamefont {E.}~\bibnamefont
  {Farhi}}, \bibinfo {author} {\bibfnamefont {J.}~\bibnamefont {Goldstone}}, \
  and\ \bibinfo {author} {\bibfnamefont {S.}~\bibnamefont {Gutmann}},\
  }\href@noop {} {\bibfield  {journal} {\bibinfo  {journal} {arXiv:1411.4028}\
  } (\bibinfo {year} {2014})}\BibitemShut {NoStop}%
\bibitem [{\citenamefont {Das}\ and\ \citenamefont
  {Chakrabarti}(2008)}]{Das2008}%
  \BibitemOpen
  \bibfield  {author} {\bibinfo {author} {\bibfnamefont {A.}~\bibnamefont
  {Das}}\ and\ \bibinfo {author} {\bibfnamefont {B.~K.}\ \bibnamefont
  {Chakrabarti}},\ }\href@noop {} {\bibfield  {journal} {\bibinfo  {journal}
  {Rev. Mod. Phys.}\ }\textbf {\bibinfo {volume} {80}},\ \bibinfo {pages}
  {1061} (\bibinfo {year} {2008})}\BibitemShut {NoStop}%
\bibitem [{\citenamefont {Albash}\ and\ \citenamefont
  {Lidar}(2018)}]{Albash2018}%
  \BibitemOpen
  \bibfield  {author} {\bibinfo {author} {\bibfnamefont {T.}~\bibnamefont
  {Albash}}\ and\ \bibinfo {author} {\bibfnamefont {D.~A.}\ \bibnamefont
  {Lidar}},\ }\href@noop {} {\bibfield  {journal} {\bibinfo  {journal} {Rev.
  Mod. Phys.}\ }\textbf {\bibinfo {volume} {90}},\ \bibinfo {pages} {015002}
  (\bibinfo {year} {2018})}\BibitemShut {NoStop}%
\bibitem [{\citenamefont {Hauke}\ \emph {et~al.}(2020)\citenamefont {Hauke},
  \citenamefont {Katzgraber}, \citenamefont {Lechner}, \citenamefont
  {Nishimori},\ and\ \citenamefont {Oliver}}]{Hauke2020}%
  \BibitemOpen
  \bibfield  {author} {\bibinfo {author} {\bibfnamefont {P.}~\bibnamefont
  {Hauke}}, \bibinfo {author} {\bibfnamefont {H.~G.}\ \bibnamefont
  {Katzgraber}}, \bibinfo {author} {\bibfnamefont {W.}~\bibnamefont {Lechner}},
  \bibinfo {author} {\bibfnamefont {H.}~\bibnamefont {Nishimori}}, \ and\
  \bibinfo {author} {\bibfnamefont {W.~D.}\ \bibnamefont {Oliver}},\
  }\href@noop {} {\bibfield  {journal} {\bibinfo  {journal} {Rep. Prog. Phys.}\
  }\textbf {\bibinfo {volume} {83}},\ \bibinfo {pages} {054401} (\bibinfo
  {year} {2020})}\BibitemShut {NoStop}%
\bibitem [{\citenamefont {Crooks}(2018)}]{Crooks2018}%
  \BibitemOpen
  \bibfield  {author} {\bibinfo {author} {\bibfnamefont {G.~E.}\ \bibnamefont
  {Crooks}},\ }\href@noop {} {\bibfield  {journal} {\bibinfo  {journal}
  {arXiv:1811.08419}\ } (\bibinfo {year} {2018})}\BibitemShut {NoStop}%
\bibitem [{\citenamefont {Zhou}\ \emph {et~al.}(2020)\citenamefont {Zhou},
  \citenamefont {Wang}, \citenamefont {Choi}, \citenamefont {Pichler},\ and\
  \citenamefont {Lukin}}]{Zhou2020}%
  \BibitemOpen
  \bibfield  {author} {\bibinfo {author} {\bibfnamefont {L.}~\bibnamefont
  {Zhou}}, \bibinfo {author} {\bibfnamefont {S.-T.}\ \bibnamefont {Wang}},
  \bibinfo {author} {\bibfnamefont {S.}~\bibnamefont {Choi}}, \bibinfo {author}
  {\bibfnamefont {H.}~\bibnamefont {Pichler}}, \ and\ \bibinfo {author}
  {\bibfnamefont {M.~D.}\ \bibnamefont {Lukin}},\ }\href@noop {} {\bibfield
  {journal} {\bibinfo  {journal} {Phys. Rev. X}\ }\textbf {\bibinfo {volume}
  {10}},\ \bibinfo {pages} {021067} (\bibinfo {year} {2020})}\BibitemShut
  {NoStop}%
\bibitem [{\citenamefont {Hastings}(2019)}]{Hastings2019}%
  \BibitemOpen
  \bibfield  {author} {\bibinfo {author} {\bibfnamefont {M.~B.}\ \bibnamefont
  {Hastings}},\ }\href@noop {} {\bibfield  {journal} {\bibinfo  {journal}
  {arXiv:1905.07047}\ } (\bibinfo {year} {2019})}\BibitemShut {NoStop}%
\bibitem [{\citenamefont {Bravyi}\ \emph {et~al.}(2020)\citenamefont {Bravyi},
  \citenamefont {Kliesch}, \citenamefont {Koenig},\ and\ \citenamefont
  {Tang}}]{Bravyi2020}%
  \BibitemOpen
  \bibfield  {author} {\bibinfo {author} {\bibfnamefont {S.}~\bibnamefont
  {Bravyi}}, \bibinfo {author} {\bibfnamefont {A.}~\bibnamefont {Kliesch}},
  \bibinfo {author} {\bibfnamefont {R.}~\bibnamefont {Koenig}}, \ and\ \bibinfo
  {author} {\bibfnamefont {E.}~\bibnamefont {Tang}},\ }\href@noop {} {\bibfield
   {journal} {\bibinfo  {journal} {Phys. Rev. Lett.}\ }\textbf {\bibinfo
  {volume} {125}},\ \bibinfo {pages} {260505} (\bibinfo {year}
  {2020})}\BibitemShut {NoStop}%
\bibitem [{\citenamefont {Farhi}\ \emph {et~al.}(2017)\citenamefont {Farhi},
  \citenamefont {Goldstone}, \citenamefont {Gutmann},\ and\ \citenamefont
  {Neven}}]{Farhi2017}%
  \BibitemOpen
  \bibfield  {author} {\bibinfo {author} {\bibfnamefont {E.}~\bibnamefont
  {Farhi}}, \bibinfo {author} {\bibfnamefont {J.}~\bibnamefont {Goldstone}},
  \bibinfo {author} {\bibfnamefont {S.}~\bibnamefont {Gutmann}}, \ and\
  \bibinfo {author} {\bibfnamefont {H.}~\bibnamefont {Neven}},\ }\href@noop {}
  {\bibfield  {journal} {\bibinfo  {journal} {arXiv:1703.06199}\ } (\bibinfo
  {year} {2017})}\BibitemShut {NoStop}%
\bibitem [{\citenamefont {Zhu}\ \emph {et~al.}(2020)\citenamefont {Zhu},
  \citenamefont {Tang}, \citenamefont {Barron}, \citenamefont {Mayhall},
  \citenamefont {Barnes},\ and\ \citenamefont {Economou}}]{Zhu2017}%
  \BibitemOpen
  \bibfield  {author} {\bibinfo {author} {\bibfnamefont {L.}~\bibnamefont
  {Zhu}}, \bibinfo {author} {\bibfnamefont {H.~L.}\ \bibnamefont {Tang}},
  \bibinfo {author} {\bibfnamefont {G.~S.}\ \bibnamefont {Barron}}, \bibinfo
  {author} {\bibfnamefont {N.~J.}\ \bibnamefont {Mayhall}}, \bibinfo {author}
  {\bibfnamefont {E.}~\bibnamefont {Barnes}}, \ and\ \bibinfo {author}
  {\bibfnamefont {S.~E.}\ \bibnamefont {Economou}},\ }\href@noop {} {\bibfield
  {journal} {\bibinfo  {journal} {arXiv:2005.10258}\ } (\bibinfo {year}
  {2020})}\BibitemShut {NoStop}%
\bibitem [{\citenamefont {B\"{a}rtschi}\ and\ \citenamefont
  {Eidenbenz}(2020)}]{Bartschi2020}%
  \BibitemOpen
  \bibfield  {author} {\bibinfo {author} {\bibfnamefont {A.}~\bibnamefont
  {B\"{a}rtschi}}\ and\ \bibinfo {author} {\bibfnamefont {S.}~\bibnamefont
  {Eidenbenz}},\ }\href@noop {} {\bibfield  {journal} {\bibinfo  {journal}
  {2020 IEEE International Conference on Quantum Computing and Engineering
  (QCE)}\ ,\ \bibinfo {pages} {72}} (\bibinfo {year} {2020})}\BibitemShut
  {NoStop}%
\bibitem [{\citenamefont {Hadfield}\ \emph {et~al.}(2019)\citenamefont
  {Hadfield}, \citenamefont {Wang}, \citenamefont {O’Gorman}, \citenamefont
  {Rieffel}, \citenamefont {Venturelli},\ and\ \citenamefont
  {Biswas}}]{Hadfield2019}%
  \BibitemOpen
  \bibfield  {author} {\bibinfo {author} {\bibfnamefont {S.}~\bibnamefont
  {Hadfield}}, \bibinfo {author} {\bibfnamefont {Z.}~\bibnamefont {Wang}},
  \bibinfo {author} {\bibfnamefont {B.}~\bibnamefont {O’Gorman}}, \bibinfo
  {author} {\bibfnamefont {E.~G.}\ \bibnamefont {Rieffel}}, \bibinfo {author}
  {\bibfnamefont {D.}~\bibnamefont {Venturelli}}, \ and\ \bibinfo {author}
  {\bibfnamefont {R.}~\bibnamefont {Biswas}},\ }\href@noop {} {\bibfield
  {journal} {\bibinfo  {journal} {Algorithms}\ }\textbf {\bibinfo {volume}
  {12}},\ \bibinfo {pages} {34} (\bibinfo {year} {2019})}\BibitemShut {NoStop}%
\bibitem [{\citenamefont {Wang}\ \emph {et~al.}(2020)\citenamefont {Wang},
  \citenamefont {Rubin}, \citenamefont {Dominy},\ and\ \citenamefont
  {Rieffel}}]{Wang2020}%
  \BibitemOpen
  \bibfield  {author} {\bibinfo {author} {\bibfnamefont {Z.}~\bibnamefont
  {Wang}}, \bibinfo {author} {\bibfnamefont {N.~C.}\ \bibnamefont {Rubin}},
  \bibinfo {author} {\bibfnamefont {J.~M.}\ \bibnamefont {Dominy}}, \ and\
  \bibinfo {author} {\bibfnamefont {E.~G.}\ \bibnamefont {Rieffel}},\
  }\href@noop {} {\bibfield  {journal} {\bibinfo  {journal} {Phys. Rev. A}\
  }\textbf {\bibinfo {volume} {101}},\ \bibinfo {pages} {012320} (\bibinfo
  {year} {2020})}\BibitemShut {NoStop}%
\bibitem [{\citenamefont {Egger}\ \emph {et~al.}(2021)\citenamefont {Egger},
  \citenamefont {Mare\v{c}ek},\ and\ \citenamefont {Woerner}}]{Egger2021}%
  \BibitemOpen
  \bibfield  {author} {\bibinfo {author} {\bibfnamefont {D.~J.}\ \bibnamefont
  {Egger}}, \bibinfo {author} {\bibfnamefont {J.}~\bibnamefont {Mare\v{c}ek}},
  \ and\ \bibinfo {author} {\bibfnamefont {S.}~\bibnamefont {Woerner}},\
  }\href@noop {} {\bibfield  {journal} {\bibinfo  {journal} {Quantum}\ }\textbf
  {\bibinfo {volume} {5}},\ \bibinfo {pages} {479} (\bibinfo {year}
  {2021})}\BibitemShut {NoStop}%
\bibitem [{\citenamefont {Tate}\ \emph {et~al.}(2020)\citenamefont {Tate},
  \citenamefont {Farhadi}, \citenamefont {Herold}, \citenamefont {Mohler},\
  and\ \citenamefont {Gupta}}]{Tate2020}%
  \BibitemOpen
  \bibfield  {author} {\bibinfo {author} {\bibfnamefont {R.}~\bibnamefont
  {Tate}}, \bibinfo {author} {\bibfnamefont {M.}~\bibnamefont {Farhadi}},
  \bibinfo {author} {\bibfnamefont {C.}~\bibnamefont {Herold}}, \bibinfo
  {author} {\bibfnamefont {G.}~\bibnamefont {Mohler}}, \ and\ \bibinfo {author}
  {\bibfnamefont {S.}~\bibnamefont {Gupta}},\ }\href@noop {} {\bibfield
  {journal} {\bibinfo  {journal} {arXiv:2010.14021}\ } (\bibinfo {year}
  {2020})}\BibitemShut {NoStop}%
\bibitem [{\citenamefont {Gra\ss}(2019)}]{Grass2019}%
  \BibitemOpen
  \bibfield  {author} {\bibinfo {author} {\bibfnamefont {T.}~\bibnamefont
  {Gra\ss}},\ }\href@noop {} {\bibfield  {journal} {\bibinfo  {journal} {Phys.
  Rev. Lett.}\ }\textbf {\bibinfo {volume} {123}},\ \bibinfo {pages} {120501}
  (\bibinfo {year} {2019})}\BibitemShut {NoStop}%
\bibitem [{\citenamefont {Perdomo-Ortiz}\ \emph {et~al.}(2011)\citenamefont
  {Perdomo-Ortiz}, \citenamefont {Venegas-Andraca},\ and\ \citenamefont
  {Aspuru-Guzik}}]{Perdomo-Ortiz2011}%
  \BibitemOpen
  \bibfield  {author} {\bibinfo {author} {\bibfnamefont {A.}~\bibnamefont
  {Perdomo-Ortiz}}, \bibinfo {author} {\bibfnamefont {S.~E.}\ \bibnamefont
  {Venegas-Andraca}}, \ and\ \bibinfo {author} {\bibfnamefont {A.}~\bibnamefont
  {Aspuru-Guzik}},\ }\href@noop {} {\bibfield  {journal} {\bibinfo  {journal}
  {Quantum Inf. Proc.}\ }\textbf {\bibinfo {volume} {10}},\ \bibinfo {pages}
  {33} (\bibinfo {year} {2011})}\BibitemShut {NoStop}%
\bibitem [{\citenamefont {Suzuki}\ \emph {et~al.}(2021)\citenamefont {Suzuki},
  \citenamefont {Kawase}, \citenamefont {Masumura}, \citenamefont {Hiraga},
  \citenamefont {Nakadai}, \citenamefont {Chen}, \citenamefont {Nakanishi},
  \citenamefont {Mitarai}, \citenamefont {Imai}, \citenamefont {Tamiya},
  \citenamefont {Yamamoto}, \citenamefont {Yan}, \citenamefont {Kawakubo},
  \citenamefont {Nakagawa}, \citenamefont {Ibe}, \citenamefont {Zhang},
  \citenamefont {Yamashita}, \citenamefont {Yoshimura}, \citenamefont
  {Hayashi},\ and\ \citenamefont {Fujii}}]{Suzuki2020}%
  \BibitemOpen
  \bibfield  {author} {\bibinfo {author} {\bibfnamefont {Y.}~\bibnamefont
  {Suzuki}}, \bibinfo {author} {\bibfnamefont {Y.}~\bibnamefont {Kawase}},
  \bibinfo {author} {\bibfnamefont {Y.}~\bibnamefont {Masumura}}, \bibinfo
  {author} {\bibfnamefont {Y.}~\bibnamefont {Hiraga}}, \bibinfo {author}
  {\bibfnamefont {M.}~\bibnamefont {Nakadai}}, \bibinfo {author} {\bibfnamefont
  {J.}~\bibnamefont {Chen}}, \bibinfo {author} {\bibfnamefont {K.~M.}\
  \bibnamefont {Nakanishi}}, \bibinfo {author} {\bibfnamefont {K.}~\bibnamefont
  {Mitarai}}, \bibinfo {author} {\bibfnamefont {R.}~\bibnamefont {Imai}},
  \bibinfo {author} {\bibfnamefont {S.}~\bibnamefont {Tamiya}}, \bibinfo
  {author} {\bibfnamefont {T.}~\bibnamefont {Yamamoto}}, \bibinfo {author}
  {\bibfnamefont {T.}~\bibnamefont {Yan}}, \bibinfo {author} {\bibfnamefont
  {T.}~\bibnamefont {Kawakubo}}, \bibinfo {author} {\bibfnamefont {Y.~O.}\
  \bibnamefont {Nakagawa}}, \bibinfo {author} {\bibfnamefont {Y.}~\bibnamefont
  {Ibe}}, \bibinfo {author} {\bibfnamefont {Y.}~\bibnamefont {Zhang}}, \bibinfo
  {author} {\bibfnamefont {H.}~\bibnamefont {Yamashita}}, \bibinfo {author}
  {\bibfnamefont {H.}~\bibnamefont {Yoshimura}}, \bibinfo {author}
  {\bibfnamefont {A.}~\bibnamefont {Hayashi}}, \ and\ \bibinfo {author}
  {\bibfnamefont {K.}~\bibnamefont {Fujii}},\ }\href@noop {} {\bibfield
  {journal} {\bibinfo  {journal} {Quantum}\ }\textbf {\bibinfo {volume} {5}},\
  \bibinfo {pages} {559} (\bibinfo {year} {2021})}\BibitemShut {NoStop}%
\bibitem [{\citenamefont {Goemans}\ and\ \citenamefont
  {Williamson}(1995)}]{Goemans1995}%
  \BibitemOpen
  \bibfield  {author} {\bibinfo {author} {\bibfnamefont {M.~X.}\ \bibnamefont
  {Goemans}}\ and\ \bibinfo {author} {\bibfnamefont {D.~P.}\ \bibnamefont
  {Williamson}},\ }\href@noop {} {\bibfield  {journal} {\bibinfo  {journal} {J.
  ACM}\ }\textbf {\bibinfo {volume} {42}},\ \bibinfo {pages} {1115} (\bibinfo
  {year} {1995})}\BibitemShut {NoStop}%
\end{thebibliography}%

\end{document}